\documentclass[a4paper,11pt]{article}
\usepackage{pos}

\title{Southern African Large Telescope Spectroscopy of BL Lacs for the CTA project}
 \ShortTitle{SALT Spectroscopy of BL~Lacs for CTA}

\author*[a]{E. Kasai}
\author[b]{P. Goldoni}
\author[a,c]{M. Backes}
\author[d]{G. Cotter}
\author[b]{S. Pita}
\author[e]{C. Boisson}
\author[f]{D. A. Williams}
\author[g]{F. D'Ammando}
\author[h]{E. Lindfors}
\author[i]{U. Barres de Almeida}
\author[j]{W. Max-Moerbeck}
\author[j]{V. Navarro-Aranguiz}
\author[k,l]{J. Becerra-Gonzalez}
\author[f]{O. Hervet}
\author[m]{J.-P. Lenain}
\author[e]{H. Sol}
\author[n]{S. Wagner}

\affiliation[a]{Department of Physics, Chemistry \& Material Science, University of Namibia, Private Bag 13301, Windhoek, Namibia\\}
\affiliation[b]{APC, AstroParticule et Cosmologie, Universit\'{e} Paris Diderot, CNRS/IN2P3, CEA/Irfu, Observatoire de Paris, Sorbonne Paris Cit\'{e}, 10, rue Alice\\}
\affiliation[c]{Centre for Space Research, North-West University, Potchefstroom 2520, South Africa\\}
\affiliation[d]{University of Oxford, Oxford Astrophysics, Denys Wilkinson Building, Keble Road, Oxford, OX1 3RH, United Kingdom\\}
\affiliation[e]{LUTH, Observatoire de Paris,  PSL Research University, CNRS,Universit\'{e} Paris Diderot, Meudon, France\\}
\affiliation[f]{Santa Cruz Institute for Particle Physics and Department of Physics, University of California, Santa Cruz, Santa Cruz, CA\\}
\affiliation[g]{INAF - Istituto di Radioastronomia, Via Gobetti 101, I-40129 Bologna, Italy\\}
\affiliation[h]{Finnish Centre for Astronomy with ESO (FINCA), Quantum, Vesilinnantie 5, FI-20014, University of Turku, Finland\\}
\affiliation[i]{Centro Brasileiro de Pesquisas Físicas (CBPF), Rua Dr. Xavier Sigaud 150, 22290-180 Rio de Janeiro, Brazil\\}
\affiliation[j]{Departamento de Astronomía, Universidad de Chile, Camino El Observatorio 1515, Las Condes, Santiago, Chile\\}
\affiliation[k]{Universidad de La Laguna (ULL), Departamento de Astrof\'isica, E-38206 La Laguna, Tenerife, Spain\\}
\affiliation[l]{Instituto de Astrof\'isica de Canarias (IAC), E-38200 La Laguna, Tenerife, Spain\\}
\affiliation[m]{Sorbonne Universit\'{e}, Universit\'{e} Paris Diderot, Sorbonne Paris Cit\'{e}, CNRS/IN2P3, Laboratoire de Physique Nucl\'{e}aire et de Hautes Energies, LPNHE, 4 Place Jussieu, F-75252 Paris, France\\}
\affiliation[n]{Landessternwarte, Universit\"at Heidelberg, K\"onigstuhl 12, D 69117 Heidelberg, Germany\\}

\forColl{CTA} 

\emailAdd{ekasai@unam.na}

\abstract{In the last two decades, very-high-energy gamma-ray astronomy has reached maturity: over 200 sources have been detected, both Galactic and extragalactic, by ground-based experiments. At present, Active Galactic Nuclei (AGN) make up about 40\% of the more than 200 sources detected at very high energies with ground-based telescopes, the majority of which are blazars, i.e. their jets are closely aligned with the line of sight to Earth and three quarters of which are classified as high-frequency peaked BL Lac objects. One challenge to studies of the cosmological evolution of BL Lacs is the difficulty of obtaining redshifts from their nearly featureless, continuum-dominated spectra. It is expected that a significant fraction of the AGN to be detected with the future Cherenkov Telescope Array (CTA) observatory will have no spectroscopic redshifts, compromising the reliability of BL Lac population studies, particularly of their cosmic evolution. We started an effort in 2019 to measure the redshifts of a large fraction of the AGN that are likely to be detected with CTA, using the Southern African Large Telescope (SALT). In this contribution, we present two results from an on-going SALT program focused on the determination of BL Lac object redshifts that will be relevant for the CTA observatory.}

\FullConference{37$^{\rm{th}}$ International Cosmic Ray Conference (ICRC 2021)\\
		July 12th -- 23rd, 2021\\
		Online -- Berlin, Germany}

\begin{document}
\maketitle

\section{Introduction}
The last two decades have seen the emergence of a new window on the Universe: very-high-energy (VHE, E >100 GeV)
gamma-ray astronomy. Thanks to three major Imaging Air Cherenkov Telescope (IACT) ground based experiments -
H.E.S.S. in the Southern hemisphere, MAGIC and VERITAS in the Northern hemisphere - over 200 sources
have been detected, both Galactic and extragalactic\footnote{http://tevcat.uchicago.edu/}. Since 2008, the Large Area 
Telescope (LAT) onboard the {\em Fermi Gamma-ray Space Telescope} - which is very sensitive up to a few tens of 
GeV - provides complementary detections at lower energies. In the coming
years, the future CTA observatory will start operations with lower
energy threshold for VHE gamma-ray detections down to a few tens of GeV with roughly an order of magnitude 
flux sensitivity improvement compared to the current-generation of IACTs.
VHE observations of active galaxies harbouring super-massive black holes and ejecting relativistic outflows
represent a unique tool to probe the physics of extreme environments, including accretion physics, jet formation,
interaction of the black-hole magnetosphere with the accretion disk corona, relativistic interaction processes and
general relativity. The same observations also allow us to characterise the evolution and differentiation (diversity,
environmental impact, feedback within the host galaxy) of some of the brightest cosmic sources through space
and time. 

The use of AGN as beacons provides insights into the cosmological evolution
of star and galaxy formation through constraints on photon fields and magnetic fields along the line of sight.
AGN are known to emit variable radiation across the entire electromagnetic spectrum up to multi-TeV energies,
with fluctuations on time-scales from several years down to a few minutes. Apart from four nearby
radio-galaxies, all VHE AGN are blazars, i.e. their jets are closely aligned with the line of sight to Earth. Three
quarters of blazars are classified as high-frequency peaked BL Lac objects but there are also a few VHE blazars of other
classes: flat-spectrum radio quasars, low and intermediate frequency peaked BL Lac objects and a newly defined
class of ultra-high-frequency peaked BL~Lac~objects, with spectral peaks above 1 TeV. The highest redshift
of this sample of VHE detected sources is $\sim$0.95 and there is some evidence of the detection of photons above
100 GeV for redshifts as large as 1.1 \cite{1}, but the majority have redshifts lower than 0.2.

The currently-known population of VHE AGN is still very limited with respect to the coverage of different classes and redshifts. 
Beyond the study of VHE BL Lac objects, their detection at different redshifts is also a valuable
tool to put constraints on the density of the extragalactic background light (EBL, see \cite{2}, \cite{3}, and \cite{4}) because 
absorption is due to pair production in the interactions between VHE and EBL photons. This EBL radiation includes the UV-optical emission of all the
stars and galaxies since the end of the cosmic dark ages and its reprocessing by dust in the near infrared. It
therefore carries valuable information about the evolution of matter in the Universe, and also plays a role as an
absorber for gamma-rays.

One challenge to studies of the cosmological evolution of BL Lacs is the difficulty of obtaining redshifts from their nearly 
featureless, continuum-dominated optical spectra. Indeed, many of the early studies using X-ray or radio-selected samples had 
highly incomplete redshift measurements, even though the samples were confined to relatively bright sources.
Uncertainty in extrapolating from the measured set of redshifts complicated
population interpretations. This difficulty has not been solved yet and it also plagues present-day
samples of BL Lacs \cite{5}. As a consequence, it is expected that a significant
fraction of the AGN detected with CTA (more than 50\% of the AGN from the {\em Third Catalogue of Hard Fermi-LAT Sources}, \cite{6}),
especially BL Lac objects, will have no spectroscopic redshifts, thus
strongly compromising studies of the BL Lac population and of its cosmic evolution. The problem is more acute
at redshifts $z > 0.3$, where known VHE sources currently become sparse. Indeed detection of TeV photons
from sources at $z > 0.3$ implies probing very high optical depths and have led to suggestions of exotic
non-standard model physics like photon-axion coupling. True spectroscopic redshifts are
thus the only way to obtain uniform population studies and constrain these theories. It is therefore of great importance to measure the redshifts of a large 
fraction of the AGN sources that are likely to be detected with CTA. 

In 2019, we proposed to observe three BL Lacs as a pilot project with the Southern African Large Telescope (SALT) to evaluate its spectroscopic capabilities in measuring redshifts. 
With an 11 m primary mirror diameter, SALT is the largest single-mirror optical telescope in the Southern Hemisphere\footnote{https://www.salt.ac.za/} and is located at the 
Sutherland Observatory in South Africa, operated by the South African Astronomical Observatory\footnote{https://www.saao.ac.za/}. SALT observed high signal-to-noise (S/N) spectra 
from which we successfully determined three redshifts. Since then, SALT observations have been on-going. 

In this contribution, we present an overview of the SALT redshift determination program, providing 
specific results of two of the first sources observed in the program for which redshifts where successfully measured and have been published in our recent 
paper \cite{7}, along with more than ten other results from other spectroscopic follow-up telescopes in the program.

The layout of the contribution is as follows: we present observations and data reduction in Section 2, redshift determination in Section 3, results in 
Section 4 and discussion and conclusions in Section 5. The calculations we perform assume a cosmology 
with $\Omega_{\rm M}$ = 0.27, $\Omega_{\rm \Lambda}$ = 0.73 and \newline$H_0$ = 70 km s$^{-1}$ Mpc$^{-1}$, and the AB system for magnitudes.  

\section{Observations and data reduction}
From November 2019 to June 2021, we have observed a total of twenty-one (21) BL Lacs, conducted with the Robert Stobie Spectrograph (RSS,~\cite{8}) -- SALT's 
main workhorse instrument that has a wide range of capabilities. We use the RSS in long slit mode with a slit width of 2" and the PG0900 grating in first-order. This configuration results in a spectral range of roughly 4500 to 7500 \AA~ with resolution $\lambda/\Delta\lambda~\sim1000$ and a throughput of more than 20 \% and \cite{9}. We use PySALT \cite{10} to perform data reduction -- correcting for cross-talk, bias, gain and flat fielding the frames -- and the standard IRAF~\cite{11} routines for wavelength calibration. Cosmic-ray cleaning and flux calibration are performed using algorithms in the Interactive Data Language\footnote{http://www.harrisgeospatial.com} software. 

Absolute flux calibration is difficult to achieve with SALT due to the telescope's moving pupil~\cite{12} and for this reason, some of our targets get observed photometrically around the same period that spectroscopic observations are conducted, including the two sources 1RXS J015658.6-530208 and 1RXS J020922.2-522920, whose observational results we present in this contribution. 1RXS J015658.6-530208 was observed on the 30$^{\mathrm{th}}$ of November 2019 in four different optical bands using the Ultraviolet/Optical Telescope (UVOT,~\citep{12}) onboard the {\em Neil Gehrels Swift Observatory}~\citep{13} and 1RXS J020922.2-522920 was observed for five nights between the end of December 2019 and the start of January 2020 using the REM Optical Slitless Spectrograph (ROSS2) at the REM telescope \citep{14}, located at an ESO observatory in Chile. 

Out of the total 21 BL Lacs observed to date, we have successfully measured redshifts for eight of them. We could not detect redshift determination spectral features in the spectra of nine of the BL Lacs, despite sufficient S/N as per our requirements and the  spectra for the remaining four of the 21 BL Lacs have lower S/N compared to our minimum requirement, necessitating further visits to those sources still.

\section{Redshift determination}
The steps involved in measuring the redshifts of our observed sources are presented in greater detail in \cite{7}. We summarise the steps in this section. The redshift determination process starts with searching for absorption or emission features in observed spectra. Whenever one such feature is found, a check is made for other features that yield the same redshift measurement result. After this step, the spectra are normalised with cubic splines and within each pixel the flux is integrated to measure each feature's total equivalent width. The error involved in such a measurement is estimated with the root of the sum of the squares of the error spectrum and considering the uncertainties of the continuum placement \citep{15}.
  
To estimate the uncertainty of a measured redshift using the above steps, two types of uncertainties are considered: (1) wavelength calibration uncertainties and (2) detected feature position uncertainties. In all our spectra, we find that from $\sim$ 4000 to $\sim$ 8000 \AA, the wavelength calibration dispersion is less than 0.5 \AA, which points to a relative precision of less than 6-12 $\times$ 10$^{-5}$.  After a redshift is measured, Gaussian functions are fitted at each feature position in the spectrum and the variance of the fits are taken to be the uncertainty. Summing such uncertainties with the wavelength calibration uncertainties results in the estimated uncertainty on a measured redshift.

\section{Results}\
 
 The results of SALT/RSS observations of 1RXS J015658.6$-$530208 and 1RXS J020922.2$-$522920 -- along with ten other sources observed by various other telescopes in the overall CTA redshift determination program -- were discussed extensively in \cite{7}. We provide a brief summary of that discussion in this section, focussing only on these two sources.
 
\begin{figure*}
   \centering
 \includegraphics[width=7.5truecm,height=6.4truecm]{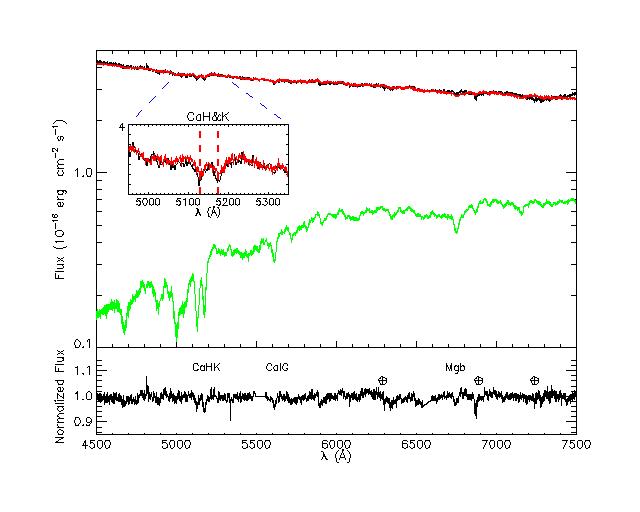} \includegraphics[width=7.5truecm,height=6.4truecm]{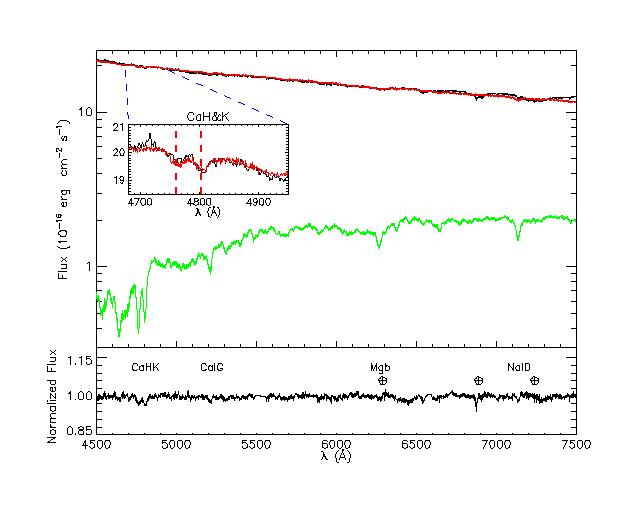} 
   \caption{SALT/RSS spectra of 1RXS J015658.6$-$530208 (left) and 1RXS J020922.2$-$522920 (right), each containing two parts. The \emph{top} parts shown in black are the flux-calibrated and telluric-corrected spectra and in red are the best fit galaxy models (where the galaxy component shown in green is assumed to be elliptical)  used by \cite{7} to estimate total blazar emission of the sources. The \emph{bottom} parts are the normalised spectra and illustrating the absorption features used for redshift determination. The symbols $\oplus$ represent atmospheric telluric absorption features. Figure from \cite{7}.}
\label{fig_spec1}
\end{figure*}
    
\subsection{1RXS J015658.6$-$530208}
 SALT/RSS observations of this target were conducted on the 24$^{\mathrm{th}}$ and 26$^{\mathrm{th}}$ of November 2019 with good transparency and average seeing of 1.2" and 1.4", respectively. The averaged spectrum from the two observations yielded a {\em S/N} = 100 and its inspection revealed a clear presence of the CaHK and CaIG features that can be seen in the bottom part of the left panel of Figure \ref{fig_spec1}. The two spectral features led to a precise redshift measurement of $z$ = 0.3043 $\pm$ 0.0004, which was confirmed by a weaker presence of the Mgb feature (also indicated in the normalised spectrum) at the same redshift.
 
\begin{figure*}[!h]
   \centering
  \includegraphics[width=7.74truecm,height=7.4truecm]{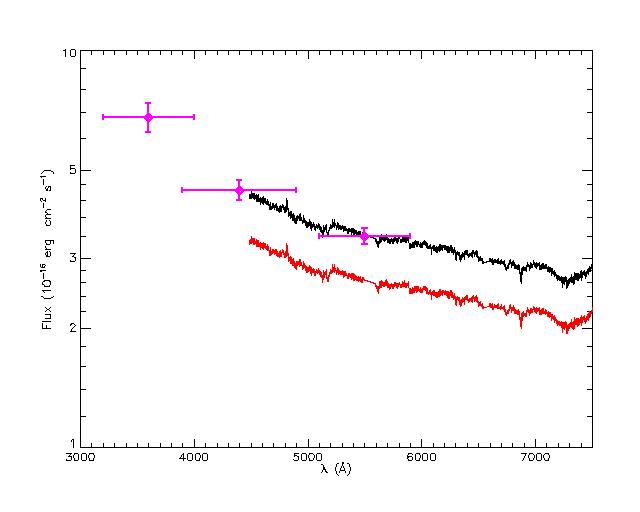}\includegraphics[width=7.74truecm,height=7.4truecm]{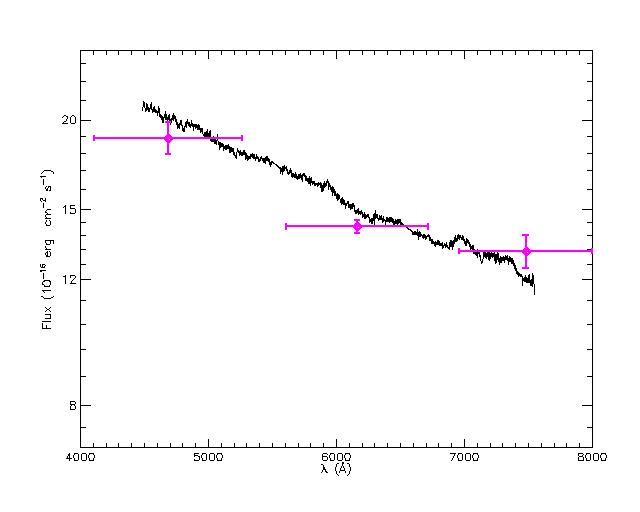}
   \caption{
In the left panel in red is the averaged SALT/RSS spectrum of 1RXSJ015658.6$-$530208 and in black is the same spectrum scaled to match the average UVOT photometry, as discussed in the text. The points in magenta show the average {\em Swift}/UVOT $u$, $b$, and $v$ photometric data points and the associated wavelength and flux error bars. In the right panel in black is the SALT/RSS spectrum of 1RXS J020922.2-522920 with the points in magenta showing the average REM/ROSS2 $g$, $r$ and $i$ photometry, found to be consistent with the spectral fluxes within at most 2$\sigma$. Figure from \cite{7}.}
 \label{SALT_photo}
    \end{figure*}

Comparison of the {\em Swift}/UVOT near-contemporaneous photometric data points to the flux-calibrated spectrum revealed that fluxes derived from the former were higher than those from the latter but the slopes were compatible, as the left panel of Figure \ref{SALT_photo} shows. 

To get more accurate spectral fluxes, a rescaling was performed to match the average UVOT photometry by multiplying the former with the value 1.3, found to be the ratio between flux from the $v$ {\em Swift}/UVOT filter (fully contained in the spectral range of the source, see left panel of Figure~\ref{SALT_photo}) and the spectral flux within the corresponding spectral $v$ filter range. 

 \subsection{1RXS J020922.2$-$522920}
 SALT/RSS observation of this target was conducted on the 28$^{\mathrm{th}}$ of December 2019 with good transparency and an average seeing of 1.4" and the resulting spectrum yielded a {\em S/N} = 160.  Upon inspection, a CaHK feature marked on the normalised spectrum in the bottom part of the right panel of Figure~\ref{fig_spec1} was identified and used to measure the redshift. A weaker presence of CaIG, Mgb and NaID features, also indicated on the normalised spectrum, confirmed the determined redshift value $z$ = 0.2110 $\pm$ 0.0002.

Comparison of the average fluxes derived from the REM/ROSS2 photometry show consistency with the SALT/RSS spectral fluxes to within 0.1 magnitudes (see right panel of Figure \ref{SALT_photo}).
  
\section{Discussion and conclusions}
As a 10m-class telescope, SALT allows reasonably short exposures to get good S/N, which is a great advantage as longer exposures result in more cosmic ray hits, which makes the data reduction process more challenging. Over 90\% of the spectra for all the 21 sources taken with SALT/RSS reached our target S/N range of 50-150, making SALT one of the best tools we have in our quest for {\em successful measurements of redshifts} for sources in our sample. The S/N of the spectra for both 1RXS J015658.6$-$530208 and 1RXS J020922.2$-$522920 were well within our target S/N interval for the former and the latter exceeded the maximum value of the interval, an aspect that is crucial to the work of the CTA Redshift Determination Group (herein CRDG). 

While the above is true, we are also cognisant of the fact that not all spectra with S/N falling within our required interval result in successful redshift measurements as the necessary spectral features are undetectable in some of such spectra. For such sources, CRDG is currently actively working towards obtaining Target of Opportunity programs on telescopes, including SALT, and observe them when in their optical low states, with a potential to result in a higher chance of detecting redshift determination features.

To conclude, SALT redshift determination program for the CTA project is ongoing. We have an approved multi-semester observing proposal on SALT running until the end of 2022, subject to approval of our semesterly performance reports by the SALT time allocation committee to continue observations in six-month intervals, in which we observe a minimum of five or six targets depending on their brightnesses.

\section*{Acknowledgements}
We gratefully acknowledge financial support from the agencies and organisations listed here: http://www.cta-observatory.org/consortium$\_$acknowledgments.

\clearpage
\section*{Full Authors List: \Coll\ Consortium}

\scriptsize
\noindent
\parindent 0pt 

H.~Abdalla\textsuperscript{1}, H.~Abe\textsuperscript{2},
S.~Abe\textsuperscript{2}, A.~Abusleme\textsuperscript{3},
F.~Acero\textsuperscript{4}, A.~Acharyya\textsuperscript{5}, V.~Acín
Portella\textsuperscript{6}, K.~Ackley\textsuperscript{7},
R.~Adam\textsuperscript{8}, C.~Adams\textsuperscript{9}
% ...

1 : Centre for Space Research, North-West University, Potchefstroom, 2520, South Africa

2 : Institute for Cosmic Ray Research, University of Tokyo, 5-1-5, Kashiwa-no-ha, Kashiwa, Chiba 277-8582, Japan

3 : Pontificia Universidad Católica de Chile, Av. Libertador Bernardo O'Higgins 340, Santiago, Chile

4 : AIM, CEA, CNRS, Université Paris-Saclay, Université Paris Diderot, Sorbonne Paris Cité, CEA Paris-Saclay, IRFU/DAp, Bat 709, Orme des Merisiers, 91191 Gif-sur-Yvette, France

5 : Centre for Advanced Instrumentation, Dept. of Physics, Durham University, South Road, Durham DH1 3LE, United Kingdom

6 : Port d'Informació Científica, Edifici D, Carrer de l'Albareda, 08193 Bellaterrra (Cerdanyola del Vallès), Spain

7 : School of Physics and Astronomy, Monash University, Melbourne, Victoria 3800, Australia

8 : Laboratoire Leprince-Ringuet, École Polytechnique (UMR 7638, CNRS/IN2P3, Institut Polytechnique de Paris), 91128 Palaiseau, France

9 : Department of Physics, Columbia University, 538 West 120th Street, New York, NY 10027, USA

% ...

\end{document}